\documentclass[12pt,onecolumn,draft]{IEEEtran}
\usepackage{color}
\usepackage{setspace}
\usepackage{comment}
\usepackage{url}
\usepackage{graphicx}
\usepackage{amssymb}
\usepackage{amsmath,amsfonts,amssymb,euscript,epsfig,psfrag,amsthm,enumerate,float,afterpage,subfigure}
\usepackage{rotate}
\usepackage{verbatim}
\usepackage{multirow}
\usepackage{wrapfig}
\usepackage{setspace}
\usepackage{subfig}

\newcommand{\squishlist}{
   \begin{list}{$\bullet$}
    { \setlength{\itemsep}{0pt}      \setlength{\parsep}{0pt}
      \setlength{\topsep}{3pt}       \setlength{\partopsep}{0pt}
      \setlength{\listparindent}{-2pt}
      \setlength{\itemindent}{-5pt}
      \setlength{\leftmargin}{1em} \setlength{\labelwidth}{0em}
      \setlength{\labelsep}{0.5em} } }

\newcommand{\squishend}{
    \end{list}  }

\newtheorem{claim}{Claim}

\def\u0{\underline{0}}

\def\eg{{\em e.g.},~}
\def\ie{{\em i.e.},~}

\def\P{{\sf P}}
\def\inn{{\rm  in}}

\def\be{\begin{eqnarray}}
\def\ee{\end{eqnarray}}
\def\beqa{\begin{eqnarray*}}
\def\eeqa{\end{eqnarray*}}

\begin{document}

\title{Network calculus for parallel 
processing}

\author{
\begin{tabular}{cc}
G.  Kesidis,
B. Urgaonkar, Y. Shan, S. Kamarava
 & J. Liebeherr\\
CSE and EE Depts & ECE Dept\\
The Pennsylvania State University  & University of Toronto\\
\{gik2,buu1,yxs182,szk234\}@psu.edu & jorg@comm.utoronto.ca
\end{tabular}
}

\maketitle

\section{Introduction}

Multi-stage parallel data processing systems are ubiquitous
in cloud computing environments. In a single
parallel processing stage, a job is partitioned into tasks
(\ie the job is ``forked" or the tasks are demultiplexed);
the tasks are then worked upon in parallel.
Within parallel processing systems,
there are often processing ``barriers'' (points of synchronization
or ``joins'') wherein \underline{all}  component tasks of a job need
to be completed before the next stage of processing of the job can
commence.
The terminus of the entire parallel processing system is typically a barrier.
Thus, the latency of a stage
(between barriers or between the exogenous job arrival point
to the first barrier) is
the greatest latency among the processing paths through it
(\ie among the tasks comprising that stage).
Google's multi-stage MapReduce
~\cite{DBLP:journals/cacm/DeanG08}
(especially its open-source implementation Apache 
Hadoop~\cite{White:2009:HDG:1717298}) 
is a very popular such framework.
However, numerous other systems exhibit similar programming patterns and our
work is relevant to them as well
\cite{DBLP:conf/osdi/YuIFBEGC08,Reiss12,google-dataflow}.

In MapReduce,  jobs arrive and  are  partitioned into tasks. 
Each task is then assigned to a mapper for initial processing. The results of
mappers are transmitted (shuffled)
to reducers. Reducers combine
the mapper results they have received and perform additional processing
(a final  stage after the reducers may simply combine their results).
The workloads of the reducer tasks may be {\em unrelated}
to those of their ``tributary" mapper tasks.
A barrier exists before each reducer (after its mapper-shuffler stage)
 and after all the reducers (after the reducer stage). 


To achieve
good interleaving of the 
principal resources consumed by the mapper (CPU/memory) and
the shuffler (network bandwidth), these stages are made to work in a 
pipelined manner wherein the shuffler transmits partial results created by the
mapper (as they are generated) rather than waiting for a mapper to 
entirely finish
its task. Of course, the shuffler must ``follow" the mapper at all times
in the sense of being able to send only what results the mapper has generated so far 
~\cite{DBLP:journals/pe/LinZWT13}. 
On the other hand, the barrier between
the shuffler stage and the reducer stage is a strict one - a reducer may
not begin any processing until all of the shuffler stage's work is done.  

Our goal is to develop a performance model for such applications.
As a first step toward this,
we consider a single parallel processing stage.
Our approach can be extended to create a model with separate 
queues for the mapper, the shuffler,
and the reducer stages (or even more stages),
but we restrict our attention to the
interaction between the shuffler and the reducer stages. 
Specifically, each processor/server (and associated
job queue) in our model represents a mapper stage.

In this paper, we focus on
the Mapper stage, where  an initial job scheduling
would be in play to achieve a ``bounded burstiness"
of the aggregate workload. We prove two claims 
using ``network calculus" for parallel processing
systems. 
We then numerically
evaluate the ``generalized" (strong) stochastic
burstiness bound (gSBB) of
a publicly disseminated
workload trace of a Facebook data-center. 


\section{Related Work}\label{sec:priorwork}


There is substantial prior work on fork-join queueing systems
particularly involving underlying Markov chains, \eg \cite{NT88}.
Our primary result is significantly simpler and somewhat more
general than those based on Markovian models of workload.
Of course, there is also an enormous literature on parallel processing
systems in general.
Typically, parallel processing systems 
employ robust load balancing to minimize
synchronization delays at the barriers. 
To this end, load balancing could proactively
estimate throughputs along the parallel processing
paths  and proportionately size the workloads from tasks fed to them. 
As an example of the many proposed reactive/dynamic mechanisms,
``straggler" (deemed excessively delayed)
tasks at barriers can be restarted or the 
entire job can be interrupted and restarted or additional
can be allocated (\eg more parallelism). 
See \cite{JianTan14,Ead14,DBLP:conf/osdi/AnanthanarayananKGSLSH10,DBLP:conf/nsdi/AnanthanarayananGSS13} 
for recent discussions on the online management specifically
for a MapReduce parallel processing system. 
Some recent work on MapReduce systems \cite{Lucent12,DBLP:journals/pe/LinZWT13}
focuses on the the pipelining between the mapper and shuffler,
the latter formulating a proactive {\em scheduling} problem that jointly
considers individual job workloads of both mapper and shuffler
(assuming the shuffler load is known {\em a priori}).

\section{Single-stage, fork-join system}\label{sec:theory}

Consider single-stage fork-join 
(parallel processing) system, 
modeled as a bank of $K$ parallel queues, with
queue-$k$ provisioned 
with service/processing capacity $s_k$. 
Let $A$ be the cumulative 
input process of work that is divided among queues
so that the $k$th queue has arrivals
$a_k$ and departures $d_k$ in such a way that
$\forall t\geq 0$,
\beqa
A(t) &=& \sum_k a_k(t).
\eeqa
Define the virtual delay processes for hypothetical departures 
from queue $k$ at time $t\geq 0$  as
\beqa
\delta_k(t) & = & t-a_k^{-1}(d_k(t)),
\eeqa
where
we define inverses $a_k^{-1}$ of non-decreasing functions
$a_k$ as continuous from the left so
 that $a_k(a_k^{-1}(v))\equiv a_k^{-1}(a_k(v))\equiv v$. 

In following  definition of the cumulative departures,
$D$, the
output is determined by the most lagging 
(straggling) queue/processor:
$\forall t\geq 0$,
\beqa
D(t)  & = & A(t-\max_k \{\delta_k(t)\})\\
& = & A\left( \min_k \{a_k^{-1}(d_k(t))\} \right).
\eeqa
Note that in the case of continuous, fluid arrivals (\eg piecewise linear $A$),
this definition of departures $D$ corresponds to 
periods of continual, possibly perpetual, barriers (synchronization times).
In the case of discrete arrivals (piecewise constant $A$ with jump discontinuities at arrival instances), then the barriers are discrete.

A queue $q$ with service $s$ has a 
(non-negative, non-decreasing) at least a  service-curve $s_{\min}$ 
(\ie $s\gg s_{\min}$) if 
for all cumulative arrivals $a$ and all time $t$
its cumulative departures 
\beqa
d(t) & \geq &  (s_{\min}*a)(t) ~:=~\min_{v\leq t} \{a(v)+s_{\min}(t-v)\}.
\eeqa

Define  the convolution ($*$) 
identity as
\beqa
u_{\infty} (t) & = & \left\{\begin{array}{ll}
0 & \mbox{if $t\leq 0$}\\
+\infty & \mbox{if $t> 0$}
\end{array}\right.\\
\eeqa
and 
\be
\lefteqn{
d_{\max,k} ~ = ~ \min\{z\geq 0 ~:~\forall x\geq 0,} & &  \nonumber \\
& &  s_{\min,k}(x) \geq (b_{\inn,k}*\Delta_z u_{\infty})(x)
=b_{\inn,k}(t-z)\} \label{dmax}
\ee
where:
\begin{itemize}
\item the delay operator
$(\Delta_d g)(t) \equiv g(t-d)$,
\item  $a_k \ll b_{\inn,k}$, \ie 
$a_k$ conforms  to the non-negative, non-decreasing
burstiness curve (traffic envelope) $b_{\inn,k}$, 
\beqa
\forall t, ~a_k(t) & \leq & (a_k*b_{\inn,k})(t),
\eeqa
\item $d_{\max,k}$ is the largest horizontal difference between $b_{\inn,k}$ 
and $s_{\min,k}$ \cite{Cruz98}.
\end{itemize}

\begin{claim}\label{deterministic-claim}
If 
the $k$th queue has at least a service curve of 
$s_{\min,k}$ and 
arrivals $a_k\ll b_{\inn,k}$,
then
for all $t\geq 0$,
\beqa
D(t) & \geq & A(t-\max_k \{d_{\max,k}\}).
\eeqa
\end{claim}

\noindent
{\bf Remark:} 
This claim simply states that the maximum delay of the whole system 
(from $A$ to $D$) is
the maximum delay among the queues.
Equivalently, the service curve from $A$ to $D$ is
at least $\Delta_d u_{\infty}$,
where $d:=\max_k d_{\max,k}$.

\IEEEproof
By Equation (\ref{dmax}), $\forall t\geq v\geq 0$ and $\forall k$,
\beqa
s_{\min,k}(t-v) & \geq & b_{\inn,k} (t-v-d_{\max,k})\\
&\geq & a_k(t-d_{\max,k})-a_k(v)
\eeqa
Thus, $\forall t\geq v\geq 0$ and $\forall k$,
\beqa
a_k(v) + s_{\min,k}(t-v) & \geq & a_k (t-d_{\max,k})\\
\Rightarrow ~~
(a_k* s_{\min,k})(t) & \geq & a_k (t-d_{\max,k})\\
\Rightarrow ~~
a_k^{-1} ((a_k* s_{\min,k})(t)) & \geq & t-d_{\max,k}\\
\eeqa
where we have used the fact that, $\forall k$,
$a_k$ are nondecreasing.
Thus,
\beqa
D(t) & = & A\left( \min_k \{a_k^{-1}(d_k(t))\} \right)\\
&\geq & A\left( \min_k \{a_k^{-1}((a_k*s_{\min,k})(t))\} \right)\\
&\geq & A\left( \min_k \{t-d_{\max,k}\} \right)\\
&= & (A*\Delta_d u_{\infty}) (t),
\eeqa
where we have used the fact that $A$ 
is nondecreasing.
\qed

We now consider a stationary stochastic model of this single-stage system.
To simplify matters, we assume the 
workload process $A$  has
strong  (``generalized")
stochastically bounded burstiness  (gSBB) \cite{gSBB09},
and leave to future work generalizations to non-stationary
settings assuming only
(weak) stochastically bounded burstiness \cite{Staro99} or
bounded log moment-generating function \cite{Chang94}.


\begin{claim}\label{stationary-claim}
In the stationary
regime at time  $t\geq 0$, if
\begin{itemize}
\item[A1] service to queue $k$, $s_k \gg s_{\min,k}$ where
\beqa
\forall v\geq 0,~~s_{\min,k}(v) &:=& v\mu_k;
\eeqa
\item[A2] 
$\forall k$, $\exists$ small $\varepsilon_k >0$  
such that $\forall v\leq t$
\beqa
\left| a_k(t)-a_k(v) - 
\frac{\mu_k}{M}(A(t)-A(v))\right| ~\leq~ \varepsilon_k 
~~\mbox{a.s.,}
\eeqa
where $M:=\sum_k \mu_k$;
\item[A3] the total arrivals have strong 
stochastically bounded burstiness
\cite{gSBB09},
\beqa
\P(\max_{v\leq t} \{A(t)-A(v)- M(t-v)\}\geq x) &  \leq &  \Phi(x),
\eeqa
where $\Phi$ decreases in $x>0$;
\end{itemize}
then $\forall x>2M\max_k \{\varepsilon_k /\mu_k\}$,
\beqa
\P(A(t)-D(t)\geq x)  & \leq & 
\Phi(x- 2M \max_k \{\varepsilon_k/\mu_k\}).
\eeqa
\end{claim}

\noindent {\em Remark:}
By A2, the mapper divides arriving work roughly proportional
to minimum allocated service resources $\mu_k$ to queue $k$, \ie strong
load balancing.

\IEEEproof
\beqa
\lefteqn{\P(A(t)-D(t)\geq x)}& & \\
 & = & \P(A(t)-A(\min_k \{a_k^{-1}(d_k(t))\})\geq x)\\
& = & \P(\min_k \{a_k^{-1}(d_k(t))\}\leq A^{-1}(A(t)-x)=:t-z)\\
& = & \P(\exists k ~\mbox{s.t.} ~d_k(t)\leq a_k(t-z))\\
& = & \P(\exists k ~\mbox{s.t.} ~a_k(t)-d_k(t)\geq a_k(t)-a_k(t-z)=: x_k),\\
& \leq & \P(\exists k ~\mbox{s.t.} ~ \max_{v\leq t}\{a_k(t)-a_k(v)-(t-v)\mu_k\}\geq x_k)
\eeqa
where we have used the fact that $A$ and the $a_k$ are
nondecreasing (cumulative arrivals) and the inequality is by A1.
Also, we have defined non-negative random variables $z$ and $x_k$ such that
$\sum_k x_k = x=A(t)-A(t-z)$.
So by using A2 then A3, we get
\beqa
\lefteqn{\P(A(t)-D(t)\geq x)}& & \\
& \leq & 
\P(\exists k ~\mbox{s.t.} ~ \max_{v\leq t} 
\{\frac{\mu_k}{M}
(A(t)-A(v))
+\varepsilon_k
-(t-v)\mu_k\} \\
& & ~~~~\geq \frac{\mu_k}{M}x-\varepsilon_k)\\
& = & \P(\exists k ~\mbox{s.t.} ~ \max_{v\leq t} \{(A(t)-A(v))
-(t-v)M\}\\
& & ~~~~~\geq x-2\frac{M}{\mu_k}\varepsilon_k)\\
& = & \P(\max_{v\leq t} \{(A(t)-A(v)) -(t-v)M\}\\
& & ~~~\geq x-2M\max_k \{\varepsilon_k/\mu_k\})\\
& \leq & \Phi(x -2M\max_k \{\frac{\varepsilon_k}{\mu_k}\}).
\eeqa
\qed

\section*{Acknowledgements}
This work was supported in part by a Cisco Systems URP gift and
NSF CAREER grant 0953541.

\bibliographystyle{plain}

\end{document}